\documentclass[final,3p,times,twocolumn]{elsarticle}
 \biboptions{comma,sort&compress}
\usepackage{here}
 \usepackage{graphicx}
  \usepackage{epsfig}

\def\nin{\noindent}
\def\beq{\begin{equation}}
\def\eeq{\end{equation}}
\def\bea{\begin{eqnarray}}
\def\eea{\end{eqnarray}}

\journal{Nuc. Phys. (Proc. Suppl.)}

\begin{document}

\begin{frontmatter}



\title{Searches for low-mass Higgs and dark bosons at BaBar}

 \author[label1]{Benjamin Oberhof\corref{cor1}}
 \author{\it \\on behalf of the BaBar collaboration.}
  \address[label1]{INFN sezione di Pisa \& Universit\'a di Pisa, \\
Polo Fibonacci - Edificio C, Largo B. Pontecorvo 3, 56125 - Pisa, Italia.}
\cortext[cor1]{Speaker}
\ead{benjamin.oberhof@pi.infn.it}



\begin{abstract}
\noindent
I present BaBar latest results for the direct search of a light CP-odd Higgs boson 
using radiative decays of the $\Upsilon(nS)$ (n=1,2,3) resonances in different final states. 
I also present the results for the search of a hidden sector gauge and Higgs bosons 
using the full BaBar datasample.

\end{abstract}


\end{frontmatter}


\section{Introduction}
\nin
Many beyond Standard Model (SM) theories account for the existence of a light 
Higgs boson. In the Next to Minimal Super Symmetric Model (NMSSM), an additional Higgs singlet 
is introduced in addition to the Minimal Super Symmetric Models (MSSM) doublets \cite{mssm} to solve the 
hierarchy problem \cite{nmssm1}-\cite{nmssm3}. The singlet and the doublet mix together to form a CP-odd 
state $A^0$ 
\begin{equation}
A^0=A_{MSSM} \cos \theta_A + A_S \sin \theta_A
\end{equation}
which mass has to be not greater than twice the bottom quark mass $m_b$ \cite{nmssm4}. 
Such a scenario is not constrained by the LEP measurements and can be explored at a low energy $e^+ e^-$ collider. 
An ideal environment for such a study are the radiative decays of the $\Upsilon(nS)$ resonances, 
i.e. $\Upsilon(nS) \rightarrow \gamma A^0$, in which $A^0$ subsequently decays to SM fermions or to invisible final states \cite{nmssm6}. 

Another scenario in which a light Higgs may arise is that of a hidden gauge sector; such theories are 
motivated by the overwhelming astrophysical evidence for the existence of dark matter \cite{dm1}-\cite{dm2}. In their simplest form 
such models introduce a new "dark" force mediated by some new "dark photon" $A'$. The dark photon is 
supposed to mix kinetically with the SM photon $A$ with a certain coupling $\epsilon$ \cite{dmt}-\cite{dmt2}. 


In these theories mass is generated via the Higgs mechanism adding one or more "dark Higges" depending on 
the specific model. Let's consider a very minimal scenario with a single dark photon and 
a single dark Higgs boson. 
If the mass of the dark Higgs and the dark photon are low enough such a scenario can be explored at low energy $e^+ e^-$ colliders 
by looking for the "Higgs-strahlung" process. In Higgs-strahlung a dark photon is created via mixing from the $e^+ e^-$ 
annihilation photon, the dark photon radiates a dark Higgs $h'$ and the $h'$ in turn decays again to 2 dark photons. 
If there are no dark sector light particles the dark photon is expected to decay to SM-fermions and the signature 
of the event is constituted by 3 pairs of oppositely charged fermions \cite{dmt3}. 



\section{Search for a light CP-odd Higgs}
\nin
The search for light CP-odd Higgs at BaBar was performed by looking to the decays of the light 
narrow $\Upsilon(nS)$ $(n=1,2,3)$ resonances; the advantage of these choice with respect to $\Upsilon(4S)$ decays, 
which represent the biggest part of BaBar data, 
is in the larger cross-section for hadron production of the light resonances as well as in the 
cleaner experimental environment they offer due to absence of backgrounds coming from $b$ quark decays. 
To search for Higgs production events two different tagging techniques were used; in the first case we look 
for the radiative decays of the $\Upsilon(2S)$ and $\Upsilon(3S)$ resonances, $\Upsilon(2,3S)\rightarrow \gamma A^0$ and to the 
subsequent decay of the $A^0$ to a pair of fermions \cite{bb1}-\cite{bb3}. 
The signature of the event is given by a monochromatic photon 
in the CM frame, and the analysis is then performed either in terms of the invariant mass of the fermion pair, if 
their energy and momentum is fully reconstructed, or in terms of the invariant mass of the photon. 
The second technique that has been used 
is to look for the decay $\Upsilon(2,3S) \rightarrow \Upsilon(1S) \pi^+ \pi^-$ followed by the radiative decay 
of the $\Upsilon(1S)$ to the CP-odd Higgs. In this case the $\Upsilon(1S)$ is reconstructed by a fit to the recoiling mass 
of the di-pion system and the analysis is performed again in terms of the invariant mass of the two fermion 
system or of the photon depending on the specific final state \cite{bb4}. 
Since the actual decay rate of the $A^0$ to SM fermions will clearly depend on the mass of the 
CP-odd Higgs different possible final states have been investigated at BaBar. 

\subsection{$\Upsilon(2,3S) \rightarrow \gamma A^0$ decays}
In these case $\mu^+ \mu^-$, $\tau^+ \tau^-$ and hadronic final states have been considered. 
For the $\mu^+ \mu^-$ channel the selection is performed requiring one photon 
with an energy greater than 200 MeV, exactly 2 tracks identified as muons and forming a common vertex; 
then an unbinned likelihood fit to the reduced mass
\begin{equation}
m_R=\sqrt{m_{\mu \mu}^2 - 4 m_{\mu}^2}
\end{equation} 
where $m_{\mu \mu}$ is the invariant mass of the di-muon system, is performed. 
No significant peaking contribution was observed in data and an UL at 90\% CL on the BR 
both for $\Upsilon(2S) \rightarrow \gamma A^0$, $A^0 \rightarrow \mu^+ \mu^-$ and 
$\Upsilon(3S) \rightarrow \gamma A^0$, $A^ \rightarrow \mu^+ \mu^-$ 
was set in the $A^0$ mass interval from 0.2 to 9.4 GeV 
respectively between $(0.26-8.6) \times 10^{-6}$ and $(0.27-5.5) \times 10^{-6}$ \cite{bb1}. 

\begin{figure}[h] 
\centerline{\includegraphics[width=7.cm]{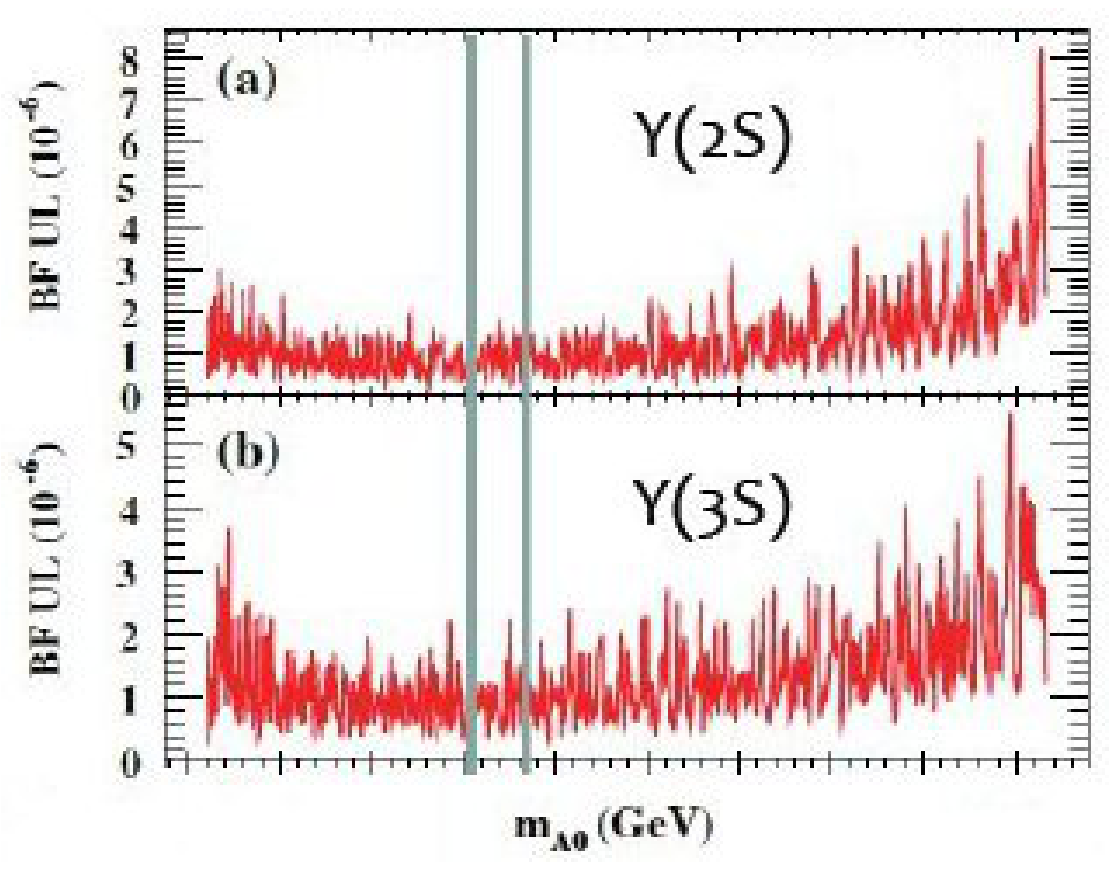}}
\caption{UL at 90\% CL for $\Upsilon(2,3S)\rightarrow \gamma A^0$, $A^0 \rightarrow \mu^+ \mu^-$ as function of the CP-odd 
Higgs mass.}
\label{mumu} 
\end{figure}

\begin{figure}[h] 
\centerline{\includegraphics[width=8.5cm]{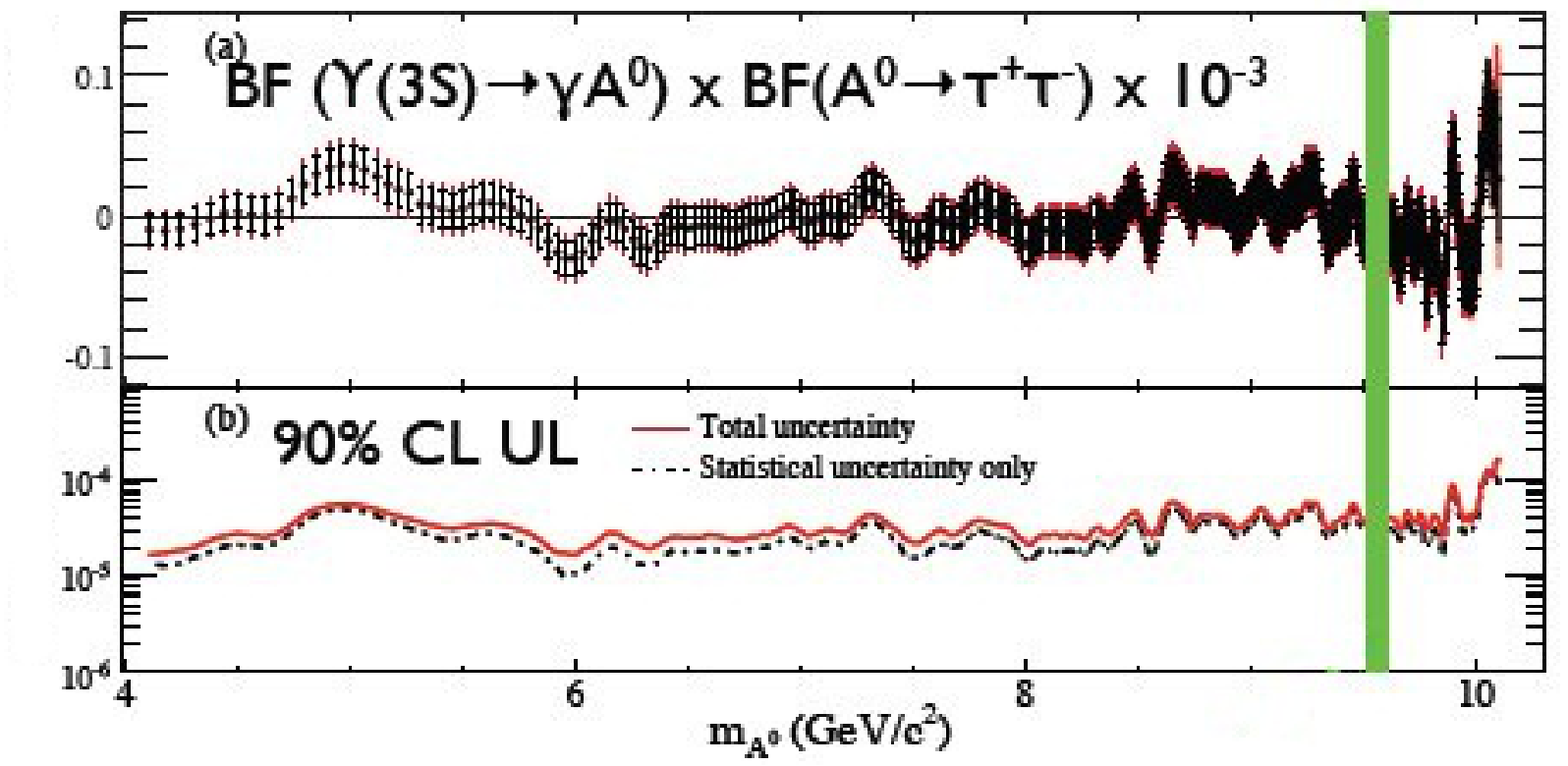}}
\caption{UL at 90\% CL for $\Upsilon(2,3S)\rightarrow \gamma A^0$, $A^0 \rightarrow \mu^+ \mu^-$ as function of the CP-odd 
Higgs mass.}
\label{tautau} 
\end{figure}

The second decay channel considered is $A^0 \rightarrow \tau^+ \tau^-$; in this case we require the two $\tau$ 
to decay leptonically, which means exactly two tracks each of them identified either as muon (anti-muon) 
or electron (positron) and a photon with an energy greater than 100 MeV. 
Since the missing energy due to the neutrinos precludes a kinematic fit to the tracks invariant mass, 
a fit to the photon energy constrained to the total CM energy has been performed. 
As can be seen from \ref{tautau} no peaking contribution was observed and this permitted to set 
an 90\% CL UL for BR($\Upsilon(2,3S) \rightarrow \gamma A^0$, $A^0 \rightarrow \tau^+ \tau^-)<(1.5-16) \times 10^{-5}$.

As last search channel in $\Upsilon(2,3S)$ radiative decays we consider hadronic decays of $A^0$. In this case we select events 
with one photon with an energy greater than 2.2 GeV ($\Upsilon(2S)$) or 2.5 GeV ($\Upsilon(3S)$) and exactly two charged tracks; 
moreover we require the full energy and momentum of the event to be reconstructed. 
For hadronic channel both CP-odd (without $K^+K^-$ and $\pi^+ \pi^-$) 
final states as well "CP-all" (i.e. without constraints) final states have been considered. 
The main backgrounds for both channels arise from $\Upsilon(nS)$ decays, namely  radiative decays to a light meson 
or non-resonant hadrons and from initial state radiation (ISR) production of a light vector meson and non-resonant hadrons. 

\begin{figure}[h] 
\centerline{\includegraphics[width=7.cm]{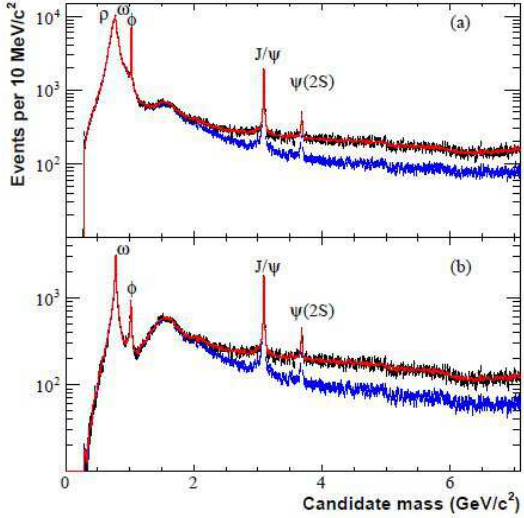}}
\caption{Candidate mass spectrum in the (a) CP-all and (b)CP-odd analyses. The top curve in each plot is the on-peak
data overlaid (in red) with the background described in the text, while the bottom curve (blue) is the scaled continuum
data. The prominent initial state radiation resonances are labeled.}
\label{had_evt} 
\end{figure}

No significant peak has been observed in data and UL at 90\% CL have been set on BR$(\Upsilon(2,3S)\rightarrow \gamma A^0$, 
$A^0 \rightarrow$ hadrons$)$ in the range $(0.1- 8) \times 10^{-5}$, assuming the same hadronic matrix element for both decays, 
depending on the $A^0$ mass \cite{bb3}.

\subsection{$\Upsilon(1S) \rightarrow \gamma A^0$ decays}
In this case the trigger is given by the $\Upsilon(2,3S) \rightarrow \Upsilon(1S) \pi^+ \pi^-$ 
decays; the recoiling mass of the two pions is used to identy the $\Upsilon(1S)$. 
BaBar made searches considering both $\mu^+ \mu^-$ and invisible 
final states of the $A^0$. For the $\mu^+ \mu^-$ final states the analysis is pretty similar to the 
$\Upsilon(2,3S)$ case: in addition to the tag-pions we require 2 charged tracks identified as muons and 
one photon with an energy $E_{\gamma}>200$ MeV, then we perform a fit on the reduced mass of the di-muon 
system and search for some peaking contribution in data. 

\begin{figure}[h] 
\centerline{\includegraphics[width=7.cm]{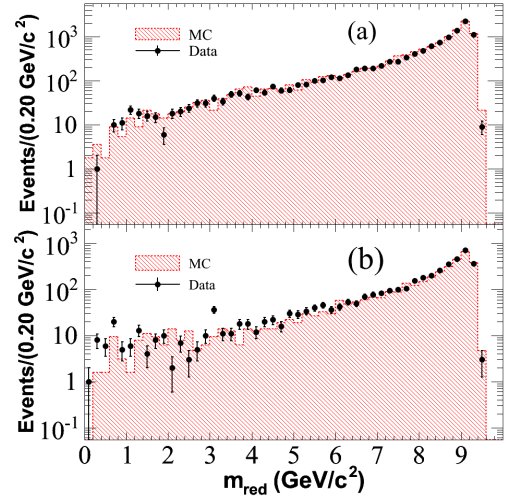}}
\caption{Distribution of the reduced mass of the $\mu^+ \mu^-$ system 
from $\Upsilon(1S)$ decays for data and background.}
\label{mumu_red_y1s} 
\end{figure}

Also in this case no significant signal was observed and we set upper limits on BR$(\Upsilon(1S) \rightarrow \gamma A^0$, 
$A^0 \rightarrow \mu^+ \mu^-)$ in the range $(0.22-10.38) \times 10^{-6}$ for the $\Upsilon(2S)+\Upsilon(3S)$
combined dataset depending on the $A^0$ mass. 

As last search for the CP-odd Higgs we consider the radiative decays of the $\Upsilon(1S)$ to invisible final states. 
This time we require the photon energy to be at least $150$ MeV, and, apart from the tagging pions, nothing 
else to be present in the event, then we perform a 2-dimensional fit to the $\pi\pi$ recoiling mass and to $E_{\gamma}$. 


In this case a peak was observed in data distribution 
corresponding to $m_{A^0}=7.58$ GeV with a significance of $2\sigma$; the probability 
to observe such a peak anywhere is $30\%$, so again we can assert that we have no evidence for a CP-odd Higgs and we 
can set an 90\% CL UL on BR$(\Upsilon(1S)\rightarrow \gamma +$Invisible$<(1.9-37)\times10^{-6}$ depending on tha $A^0$ mass
\cite{bb4}. 

\begin{figure}[h] 
\centerline{\includegraphics[width=6.cm]{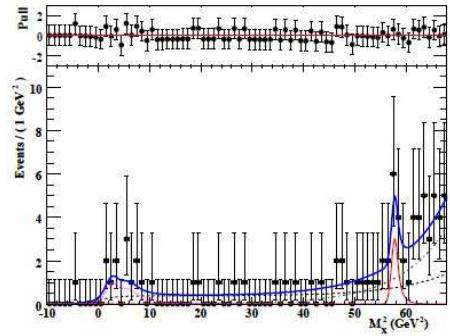}}
\caption{Projection plot from the fit with $m_{A^0} = 7.58$ GeV $M^2_X$. Overlaid is the fit 
(solid blue line), signal contribution (solid red line), continuum background (black dashed line),
radiative leptonic $\Upsilon (1S)$ decays (green dash-dotted line), 
and radiative hadronic $\Upsilon(1S)$ decays (magenta dotted line).}
\label{missing_mass} 
\end{figure}

Using tha same data we can also set an UL on the BR for the process $\Upsilon (1S) \rightarrow \gamma \chi \chi$, where 
$\chi$ is some dark matter particle candidate in the range $(0.5-24) \times 10^{-5}$ depending on $m_{\chi}$ \cite{bb4}.



\nin
\section{Search for dark bosons}
\nin

As already said in the introduction we want to search for Higgs-strahlung processes in which all three dark photons 
decay to fermion pairs. The advantage of looking to such a process lies in the fact that it is suppressed only 
by a single $\epsilon$ factor and that it has a very small background from SM processes. 
BaBar considered either final states with 6 leptons (electrons or muons) or final states with 
two leptons pairs and one $A'$ decaying to hadrons; in detail, the admitted combinations are 
$6 \mu$, $4\mu 2e$, $2\mu4e$, $6e$, $4\mu 2p$, $2\mu2e2p$, $4e2p$, $2\mu4p$, $2e4p$, $4\mu +X$, $2\mu 2e+X$. 
According to the nature of the final state two different selection criteria were used: in full reconstruction 
we require all 6 tracks to be reconstructed and to be identified either as electrons, muons or pions in accord 
to the aforementioned combinations, while in partial reconstructions we require the first pair to be either 
electrons or muons while the second have to be muons. We don't make any assumption on the third (missing) pair 
and we calculate the missing 4-momentum from the reconstructed tracks. 
After this preliminary selection we require the masses of all 3 reconstructed $A'$ to be the same and we plot 
the entries on the $m_{h'}, m_{A'}$ plane. Fore every event which passes the selection we have three different entries 
depending on which photons we assume to come from the dark Higgs decay. 

\begin{figure}[h] 
\centerline{\includegraphics[width=7.cm]{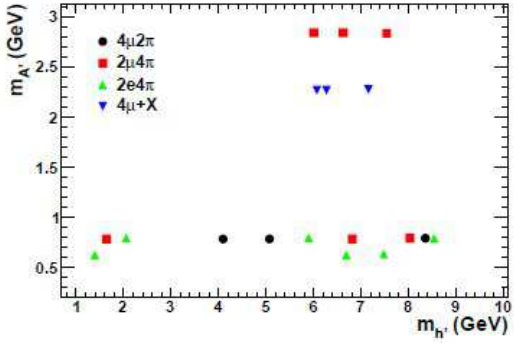}}
\caption{Candidate signal events in the $m_{h'}$, $m_{A'}$ plane for dark Higgs-strahlung.}
\label{llllll} 
\end{figure}

The full BaBar dataset ($\sim 530$ fb$^{-1}$) 
was used for this search and 6 candidate signal events passed the selection. No 
event with 6 leptons was observed. This is consistent with the background only 
hypothesis obtained from control samples. 
From this result we can set an upper limit and the cross section for Higgs-strahlung 
$\sigma (e^+ e^- \rightarrow A' h'$, $h' \rightarrow A' A') < (10-100)$ ab$^{-1}$, 
depending on the boson masses, as well as on the product of the dark coupling constant $\alpha_D$ with 
the mixing parameter $\epsilon$, $\alpha_D \epsilon^2 < 10^{-6}$ at 90\% CL.

\begin{figure}[h] 
\centerline{\includegraphics[width=7.cm]{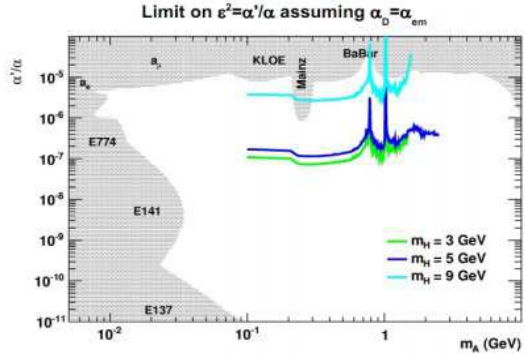}}
\caption{90\% CL UL on $\epsilon^2 = \alpha'/\alpha$ assuming $\alpha_D = \alpha_{QED}$ 
for various $m_{h'}$ values from BaBar results. 
The peaks are due to $\omega$ and $\phi$ resonances.}
\label{llllll_ul} 
\end{figure}

\section{Conclusions}
\nin
BaBar performed direct searches both for a light CP-odd Higgs as well as for dark sector 
bosons in different channels and no significant signal was observed in any of them. 
This permitted to set improved limits with respect to precedent measurements 
on the BRs for such processes and to exclude new regions in the parameter 
space of New Physics models. 
Further results from other channels are expected in the near future.















\end{document}